\documentstyle[prl,aps,multicol,epsf]{revtex}
\begin{document}
\title{New Superhard Phases for 3D  C$_{60}-$based Fullerites}
\author{E. Burgos$^1$, E. Halac$^1$, Ruben Weht$^{1,2}$, H. Bonadeo$^1$, Emilio
Artacho$^3$ and Pablo Ordej\'on$^2$}
\address{$^1$Departamento de F\'{\i }sica, CNEA, Avda. Gral. Paz 1499, (1650) San
Mart\'\i n, Pcia. de Buenos Aires, Argentina}
\address{$^2$Institut de Ci\`{e}ncia de Materials de Barcelona - CSIC, Campus de la
U.A.B.  E-08193 Bellaterra, Barcelona, Spain}
\address{$^3$Departamento de F{\'{\i }}sica de Materia Condensada C-III 
and Institute Nicol\'as Cabrera, \\ 
Universidad Aut\'{o}noma de Madrid, E-28049 Madrid, Spain.}
\maketitle

\begin{abstract}
We have explored new possible phases of 3D C$_{60}$-based fullerites using
semiempirical potentials and {\em ab-initio} density functional 
methods. We have found three closely related structures - two body centered
orthorhombic and one body centered cubic - having 52, 56 and 60
tetracoordinated atoms per molecule. These 3D polymers result in
semiconductors with bulk moduli near 300 GPa, and shear moduli around 240
GPa, which make them good candidates for new low density superhard materials.
\end{abstract}


\begin{multicols}{2}

Superhard materials are of obvious practical and theoretical interest. Among
them, diamond is still the undisputed hardest substance, and will take some
beatings. Its hardness comes from the stiffness of its tetravalent carbon
bonds, which are also the basis of the widely used diamond-like coatings.
However, natural diamond is rare and its industrial production expensive.
For this reason, much experimental and theoretical work has been devoted to
studying novel potential candidates for hard materials. For instance,
another carbon-based crystal, 3D $sp^{2}$ carbon~\cite{graph3d}, was
theoretically predicted to be even harder than diamond but never
synthesized. Very recently, a high pressure phase of silicon clathrates was
studied~\cite{clath}. Its bulk modulus resulted to be 90 GPa, very close to
that of the normal silicon cubic crystal, leading to the speculation that,
if constructed with carbon instead of silicon, it could be as hard as
diamond. A similar idea was proposed for $\beta$-C$_{3}$N$_{4}$ in relation
to $\beta$-Si$_{3}$N$_{4}$~\cite{c3n4} but because only small quantities of
the former material are available, it remains to be proven that the
substance is indeed harder than diamond. A general theoretical approach to
study hardness in any of its definitions is difficult, if not impossible, to
perform. Traditionally, it has been linked to a large bulk modulus but it
has been shown recently to correlate better with the shear modulus of the
material~\cite{teter}; diamond is the leading system in both cases, with 443
and 535 GPa, respectively.

The discovery of C$_{60}$ in 1985 as a third form of crystalline carbon~\cite
{c60} was the kick-off for one of the most stimulating fields for research
in materials science in the last decades. For the first years after its
discovery and massive production, the known solid state phases of pure C$%
_{60}$ were nearly ideal molecular crystals~\cite{c60mc}. However, from the
very beginning, there were speculations on the possibility of obtaining
crystals in which the C$_{60}$ molecules were covalently bonded to each
other, by means of high pressures and temperatures~\cite{kratch,ruoff,okif}.
Since then, several papers~\cite{manolo1,iwa,aga,moret,isa} have appeared
reporting a variety of experimental results in this line (see Ref. %
\onlinecite{sund} for an updated review of the field).

Nu\~nez-Regueiro {\em et al.}~\cite{manolo2} reported the crystal structures
of partially covalent forms of C$_{60}$. They observed a crystal containing
linear chains of polymerized fullerene molecules and also two structures,
one rhombohedral and one orthorhombic, of covalently bonded molecular planes
(2D C$_{60}$). The structures are formed by bonding through 2+2
cycloaddition reactions between neighboring molecules with a carbon
hybridization change from $sp^2$ to $sp^3$; chains and planes are bound
together by Van der Waals interactions. More recently, the realization of 3D
C$_{60}$-based fullerite
has been reported~\cite{marquez}. At least four crystal structures
were found to be compatible with the X-ray data: rhombohedral, orthorhombic
(pseudo tetragonal), tetragonal and simple cubic. Of these, the first two
are formed by stacking covalent rhombohedral and tetragonal planes. 
Blank and coworkers~\cite{blank} have reported the obtention 
of 3D fullerene solids of extremely 
low compressibility, and very recently a superhard structure has been 
also reported in detail by Chernozatonskii {\em et al.}~\cite{rusos}; the 
crystal is body centered orthorhombic, quasi fcc, and each molecule develops 
two intermolecular bonds with each of its 12 neighbors, presenting 
corrugated surfaces. From the theoretical side, Okada {\em et al.}~%
\cite{oka} have reported calculations on a possible structure for
pseudo-tetragonal (body centered orthorhombic) covalently bonded 3D C$_{60}$%
, which might be obtained by applying a uniaxial pressure of about 20 GPa to
pseudo-tetragonal polymerized layers of C$_{60}$. This phase would be
metallic with a high density of states at the Fermi level, and the C$_{60}$
molecules would be highly corrugated.

In this work we report new 3D C$_{60}$-based fullerite phases which have a
direct relation with the most probable precursors: polymerized linear chains
and orthorhombic planes, while the molecules preserve their convexity. We
will show that our proposed structures have low compressibilities and large
shear moduli and therefore are excellent candidates for hard materials.

\end{multicols}

\begin{figure}[tbp]
\epsfxsize=16.0cm\centerline{\epsffile{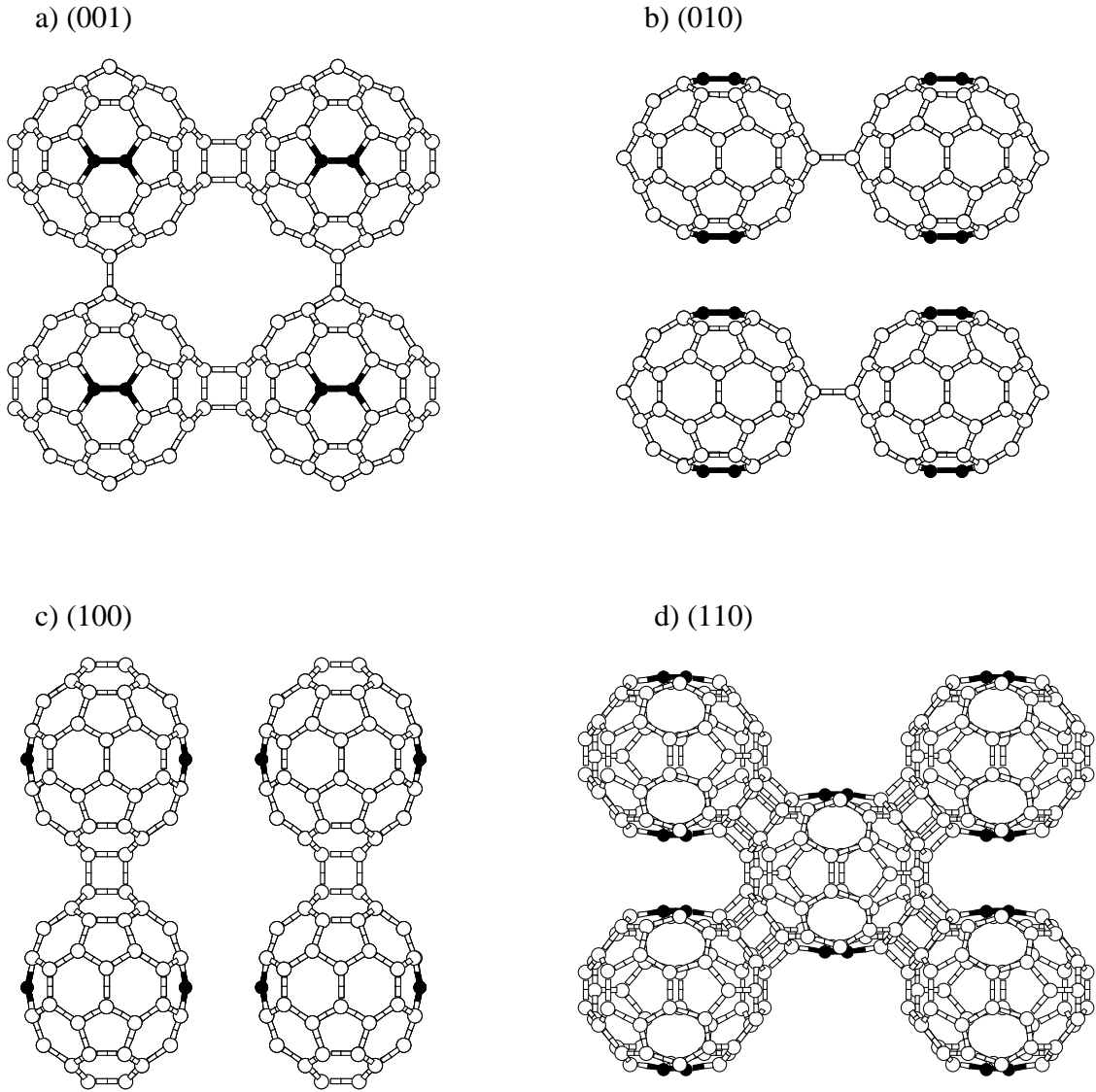}}
\caption{Views of the crystal structure of the proposed hard (56-4)
fullerene solid.
The hexagonal turrets connecting nearest neighbours molecules are
clearly seen in d). Tricoordinated atoms are shown in black.}
\label{FigXtal}
\end{figure}
 
\begin{multicols}{2}

We have proceeded as follows: an initial search of the stable or metastable
structures was done using Tersoff potentials~\cite{tersoff}, which are known
to provide a good description of carbon fullerenes~\cite{bhb1,bhb2},
including several polymeric phases~\cite{bhb3}. Once the potentially
interesting structures were found, these were used as starting points for
more detailed and accurate {\em ab-initio} calculations.
The structure, stability and electronic properties of each 
phase were obtained, allowing the complete relaxation of cell parameters and
atomic coordinates, and analyzed; all data we will give here (except for the 
calculated vibrational frequencies) are those of the final {\em ab-initio} 
results.

The {\em ab-initio} calculations were performed using a
numerical-atomic-orbital density functional (DFT) method described in detail
elsewhere~\cite{siesta1}. It has been already applied to large fullerene
molecules and nanotubes~\cite{siesta2}, and many other systems~\cite
{siesta-rev}. The calculations are done using the Generalized Gradient
Approximation for the exchange-correlation, as parametrized by Perdew, Burke
and Ernzerhof~\cite{pbe}. Core electrons are replaced by nonlocal,
normconserving pseudopotentials~\cite{tm2} factorized in the
Kleinman-Bylander form~\cite{kb}, whereas valence electrons are described
using linear combinations of pseudo-atomic orbitals. In this work we have
used a split-valence double-$\zeta $ basis set, supplemented with
polarization $d$ orbitals (DZP)~\cite{emilio}. The radial cutoffs were 4.2
and 5.0 a.u. for the $s$ and $p$ orbitals respectively. We have carried out
tests using a triple-$\zeta $ basis set, finding only minor changes in the
lattice constants (less than 0.3 \%) and elastic moduli. The charge density
was expanded in plane waves with a large cutoff of more than 150 Ry (the
exact values vary slightly, depending on the volume). With these parameters
we have obtained a bulk modulus for diamond of 430 GPa, in good agreement
with known values~\cite{teter}. The relaxation of the cell parameters
and atomic coordinates was  performed using conjugated gradient
minimizations at fixed pressure.

Starting from two experimentally observed 1D and 2D polymerized body
centered orthorhombic C$_{60}$ structures, we obtain two new metastable 3D
phases. These have 52 and 56 tetracoordinated atoms per molecule,
respectively, and we will refer to them as (52-8) and (56-4). The (56-4)
structure is obtained by applying uniaxial pressure along the stacking
direction of a 2D polymer, while for the (52-8) one the pressure is applied
in the plane perpendicular to the direction of linear chains. In those
phases, the individual molecules preserve their convexity as well as the
bonds between C$_{60}$ molecules originally present in the chains or planes
(4 and 8, respectively). In both cases, six new bonds are formed with each
of the new nearest neighbor molecules in the body centered position. Both
structures are intimately related to one proposed by O'Keeffe, based on
purely geometric arguments~\cite{okif}: it is possible to construct a body
centered cubic crystal with $T_h$ symmetry, having all carbon atoms with
coordination four, connected by bonds of equal length. We have also studied
this phase, from which, once all independent degrees of freedom are relaxed,
a (60-0) structure is obtained. From the electronic point of view, all the
new structures are semiconductors, with bands gaps larger than 2 eV. A
general view of the (56-4) structure is schematically shown in Fig. \ref
{FigXtal}, where also the few remaining {\em sp$^2$} atoms can be seen. The
(52-8) phase differs from the (56-4) in that both the (001) and the (010)
planes look like the (001) plane in Fig. \ref{FigXtal}. For the (60-0)
structure, the (001), (010) and (001) planes, which are equivalent by
symmetry, are bonded as in Fig. \ref{FigXtal}-a.

Table I shows the relevant structural data and physical parameters of the
new fullerites~\cite{position}. The specific volumes lie between those
corresponding to graphite (9 ${\rm \AA ^3}$/atom) and diamond (5.6 ${\rm \AA
^3}$/atom) but due to the covalent intermolecular bonds are much smaller
than those of Van der Waals C$_{60}$ and C$_{70}$ crystals 
($\approx $ 11.5 ${\rm \AA
^3}$/atom). The length of the bonds between tricoordinated atoms is about
1.35 ${\rm \AA }$; between tri- and tetracoordinated atoms about 1.48 ${\rm %
\AA }$; the remaining intramolecular bond lengths range from 1.48 to 1.70 $%
{\rm \AA }$. The bonds forming the hexagonal turrets which connect molecules
in the body centered position are about 1.57 ${\rm \AA }$ long, and those
connecting second neighbor molecules 1.62 and 1.66 ${\rm \AA }$. As
expected, the bond angles are grouped around the values of graphite,
diamond, and 90 degrees; the latter are associated with the turrets and
intermolecular second neighbor bonds. For (60-0), where only {\em sp$^3$}
bonds are present, we obtain bonding distances of 1.475, 1.548 and 1.56 $%
{\rm \AA }$. Regarding the intermolecular distances between first neighbor
molecules in the (110) plane ({\em i.e.}, those

\begin{figure}[tbp]
\epsfxsize=5.5cm\centerline{\epsffile{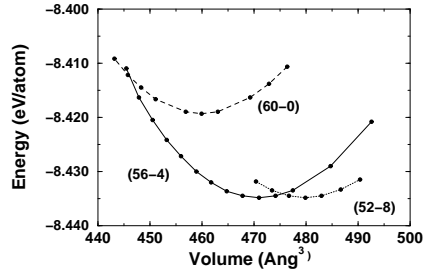}}
\caption{Energy vs. volume curves for the (56-4) (solid lines), (52-8)
(dotted lines) and (60-0) (dashed lines) structures.}
\label{FigEvsV}
\end{figure}

\noindent
corresponding to the maximum
number of covalent bonds between C$_{60}$ molecules), the typical values are
around 8.4 - 8.5 ${\rm \AA }$, which agree very well with those
experimentally found by Marquez {\em et al.}~\cite{marquez} (8.40 and 8.80 $%
{\rm \AA }$).

The number of $sp^3$ bonds suggested the interesting possibility of low
compressibility materials, and in fact our structures have bulk and shear
moduli around 300 GPa and 240 GPa, respectively. They should be compared to
the values calculated with the same method for diamond, 430 and 560 GPa.
According to the empirical relation found by Teter between hardness and
shear modulus~\cite{teter}, these values would correspond to a hardness of
about 40 Vickers, which would place them among the ten or so hardest known
systems, at the lower limit of superhard materials. It is also interesting that
the hardness expected for these phases comes together with a considerable 
lightness of the materials, since the densities are about 30 \% smaller than 
that of diamond.

Energy vs. volume curves for the new phases are shown in Fig. \ref{FigEvsV}.
It can be seen that the (56-4) and (52-8) structures have almost the same
equilibrium energy and are more stable than the (60-0) by about 1 eV. A
suggested (56-4) $\rightarrow $ (60-0) transition is found at about 14 GPa, in
which the (56-4) structure is compressed along one of its axes. We have also 
recalculated the structure of Okada {\em et al.}~\cite{oka} finding that the 
(56-4) crystal is lower in energy by almost 2 eV, and has a smaller volume. 
The structures obtained in this work (and that of
Ref.20) are metastable with respect to the corresponding precursor planes
and chains, and correspond to local minima in configuration space. However,
as seen in Fig.2, they have a large stability range, and require quite
considerable amounts of energy to be dissociated. 

\begin{table}[tb]
\caption{Unit cell parameters (${\rm \AA }$){\rm ;} specific volumes {\em v}
(${\rm \AA ^{3}/}$atom) and Bulk Moduli B$_{\text{o}}$ (GPa) for the 
new proposed structures.}
\begin{tabular}{cccccc}
Structure & {\em a} & {\em b} & {\em c} & {\em v} & B$_{\text{o}}$ \\ \hline
52-8 & 9.90 & 9.76 & 9.92 & 7.99 & 298 \\ 
56-4 & 9.76 & 9.78 & 9.88 & 7.86 & 300 \\ 
60-0 & 9.73 &  &  & 7.67 & 295
\end{tabular}
\end{table}

\begin{figure}[tbp]
\epsfxsize=6.0cm\centerline{\epsffile{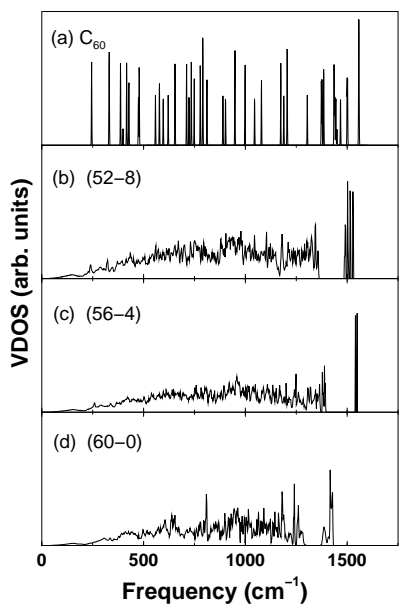}}
\caption{Vibrational densities of states for the C$_{60}$ molecule and the
(52-8), (56-4) and (60-0) structures.}
\label{Vib}
\end{figure}
             
It is interesting to note that these phases could be recognized by
vibrational spectroscopy methods. Fig. \ref{Vib} 
shows the vibrational DOS for the isolated C$_{60}$ molecule, 
and the (52-8), (56-4) and (60-0) structures, 
calculated using the potentials described in Ref.~\cite{bhb2}; the bands 
above 1500 cm$^{-1}$, which are related to graphite-like vibrations in 
C$_{60}$ (the in-plane optical frequencies of graphite lie around 
1590 cm$^{-1}$), completely disappear in the (60-0) crystal; only four isolated 
optical bands remain in the (56-4) compound, and eight in the 52-8 one.

In conclusion, we have studied new phases of 3D C$_{60}$-based fullerites which 
seem to have quite interesting features: their bulk moduli are about 70 \% 
of diamond and their shear moduli compare with those of the hardest systems 
known; the phases are extremely compact, although the crystals would be 
light compared to most other hard materials; the original convexity of the 
molecules is preserved - corrugation is a costly energy consuming process; 
and the (52-8) and (54-6) structures are directly related to polymerized 
chains and layers, which seem to be most likely precursors for 3D 
polymerized phases, and could probably be obtained at extreme, possibly
uniaxial, pressures. The present result, joined to that of the metallic
phase reported by Okada {\em et al.}~\cite{oka} and the latest experimental
data, suggest an exciting scenario for research on new fullerene based
materials.

This work was partially supported by CONICET Grant PMT-PICT 0051. R.W.
acknowledges support from Fundaci\'on Antorchas Grants No A-13622/1-103 and
A-13661/1-27. P.O and R.W. acknowledge support by Motorola PSRL. 
Part of the research has been done using the computing resources 
of CESCA and CEPBA, coordinated by C$^4$.

\end{multicols}

\end{document}